\documentclass[fleqn,usenatbib]{mnras}

\usepackage{newtxtext,newtxmath}

\usepackage[T1]{fontenc}
\usepackage{hyperref}

\DeclareRobustCommand{\VAN}[3]{#2}
\let\VANthebibliography\thebibliography
\def\thebibliography{\DeclareRobustCommand{\VAN}[3]{##3}\VANthebibliography}

\usepackage{todonotes}
\usepackage{graphicx}	
\usepackage{amsmath}	
\usepackage{float}

\title[Subcluster Mergers]{Modelling Star Cluster Formation: Mergers }

\author[Karam \& Sills]{
Jeremy Karam \thanks{E-mail: karamj2@mcmaster.ca}
and Alison Sills
\\
Department of Physics \& Astronomy, McMaster University, 1280 Main Street West, Hamilton ON, L8S 4M1, CANADA
}

\date{Accepted XXX. Received YYY; in original form ZZZ}

\pubyear{2022}

\begin{document}
\label{firstpage}
\pagerange{\pageref{firstpage}--\pageref{lastpage}}
\maketitle

\begin{abstract}

Star cluster formation in giant molecular clouds involves the local collapse of the cloud into small gas-rich subclusters, which can then subsequently collide and merge to build up the final star cluster(s). In this paper, we simulate collisions between these subclusters, using coupled smooth particle hydrodynamics for the gas and N-body dynamics for the stars. We are guided by previous radiation hydrodynamics simulations of molecular cloud collapse which provide the global properties of the colliding clusters, such as their stellar and gas masses, and their initial positions and velocities. The subclusters in the original simulation were treated as sink particles which immediately merged into a single entity after the collision. We show that the more detailed treatment provides a more complex picture. At collisional velocities above $\approx$ 10 km/s, the stellar components of the cluster do not form a monolithic cluster within 3 Myr, although the gas may do so. At lower velocities, the clusters do eventually merge but over timescales that may be longer than the time for a subsequent collision. The structure of the resultant cluster is not well-fit by any standard density distribution, and the clusters are not in equilibrium but continue to expand over our simulation time. We conclude that the simple sink particle treatment of subcluster mergers in large-scale giant molecular cloud simulations provides an upper limit on the final cluster properties. 

\end{abstract}

\begin{keywords}
star clusters: general -- stars: kinematics and dynamics -- stars: formation
\end{keywords}



\section{Introduction}

Star clusters are an integral building block of galaxies. Most stars form in clustered environments, and many of these stellar groupings disperse into the field of the galaxy over different timescales, ranging from a single crossing time to many Gyr, depending on the initial properties of the cluster. Star cluster formation mechanisms have been explored through a variety of computational simulations, many of which show that star cluster formation takes place through hierachical mergers of smaller subclusters comprised of groups of stars and gas. Simulations of cluster evolution provide insight into the timescales of these mergers as well as their effect on the components of the final star cluster. In most simulations, mergers happen quickly and cause an overall expansion of the resultant cluster's gas component (\citealt{BK2015}, \citealt{sills2018}). They also show us that this merger process takes place along dense gaseous filaments in giant molecular clouds (GMCs) and may eventually lead to the creation of a central massive cluster through multiple subcluster mergers (\citealt{Howard2018}, \citealt{li2019}, \citealt{lahen}). 

Observations support this formation mechanism as well. For example, a study of radial velocities of different stellar populations in the star cluster Westerlund 2 shows two distinct clumps on route to merging in the near future (\citealt{ziedler2021}). \citet{fujiizwart2012} also showed that the existence of runaway stars around the young cluster R136 could be explained through prompt subcluster mergers in a time frame of $3$ Myr.
As a consequence of these mergers, star cluster masses can build to $\gtrsim 10^4$M$_\odot$ in times $\lesssim 4$Myr 
. The stellar component of such young clusters has not had enough time to expel the gaseous component through stellar feedback and supernovae and should therefore be modelled as a collection of stars and gas (\citealt{pelupessy2013}).

GMC simulations, such as those performed by \citet{Howard2018} (hereafter, H18), involve regions with large density contrasts between the ambient GMC and star forming regions. Because subclusters were the most dense regions of their simulations, H18 used the sink particle prescription outlined in \citet{sink} to model them in their GMC. 
 
Each sink particle carries with it a set of parameters that describe the stellar cluster that it represents (e.g. position, velocity, total mass, and mass in stars). The sink particle size is dictated by a constant accretion radius proportional to the highest resolution obtainable by the hydrodynamics code being used. Once a merger takes place between two sink particles, the resulting sink particle will inherit the combined mass of its parent sinks; however, its accretion radius will remain unchanged. 

This sink particle approach to star cluster modelling is used to allow the hydrodynamics simulations to proceed with reasonable resolution over reasonable timescales. However, physical processes that occur on smaller scales such as gas expulsion (e.g \citealt{gb2001}, \citealt{bk2007}, \citealt{pelupessy2013}, \citealt{gasexp}) and mass segregation (\citealt{mass_seg}) cannot be followed. Cluster properties such as size and bound stellar and gas fractions have been found to change with cluster evolution (\citealt{pelupessy2013}). Understanding the internal evolution of the clusters, especially during and after cluster mergers, could have a significant impact on the final properties of the clusters formed in these simulations.

In order to explore the detailed evolution of the merger of young star-forming clusters, we model the subcluster mergers seen in the H18 simulations explicitly, including gravitational interactions between the stars and gas in each subcluster. We analyze the resultant clusters and investigate how the distribution of stars and gas have been altered by the merger process.

In section \ref{sec:setup} we discuss the initial conditions used to set up our star clusters. In section \ref{sec:rcp} we look at the resultant cluster properties of a sample merger. In section \ref{sec:suite} we generalize our results from section \ref{sec:rcp} for all our simulated mergers. In section \ref{sec:conc} we discuss our results and their implications to the general use of sink particles.

\section{Method}
\label{sec:setup}
In the following sections, we discuss the numerical methods being used in our simulations, outline the process by which we initialize our isolated clusters, and describe the initial set up of our merger simulations.

\subsection{Numerical Methods}
Our simulations are performed using the Astrophysical Multipurpose Software Environment (AMUSE) (\citealt{zwart2009}, \citealt{pelupessy2013}) which contains many codes that evolve the equations of gravity and hydrodynamics. It also allows for communication between codes, allowing us to simultaneously simulate the dynamics of the stars and hydrodynamics of the gas in the cluster. 

For our N-body code we use \texttt{hermite0} (\citealt{hermite}) and for our smoothed particle hydrodynamics (SPH) code we use \texttt{GADGET-2} (\citealt{gadget2}). For the communication scheme, we use \texttt{BRIDGE} (\citealt{bridge}) with \texttt{BHTree} (written by Jun Makino based on \citealt{tree}) as the connecting algorithm. This scheme feeds outputs from one of the solvers (i.e. positions, velocities, and masses) to the other so that the stars may react to changes in the gas properties, and the gas may react to changes in the star's properties. This communication procedure happens once every bridge timestep $t_B$, which should be sufficiently small that energy is conserved, while still allowing for resonable simulation times. For our simulations, a choice of $t_B \approx 800$yr is reasonable.

\subsection{Isolated Clusters}

We set up initial conditions for our individual clusters using parameters taken from the H18 sink particles, specifically, the mass in stars $M_{*}$ and the mass in gas $M_g$. The initial distributions of our N-body and SPH particles follow a Plummer (\citealt{plummer1911}) sphere with a scale radius chosen such that the half mass density of the cluster starts as $\rho_{hm} = 10^3M_\odot/pc^3$. This density choice is consistent with observed densities of young massive clusters (\citealt{zwart}). Given a $\rho_{hm}$, and the total mass of the cluster, we can calculate the half mass radius $r_{hm}$ which we convert to a Plummer scale radius using $a \approx 1.3r_{hm}$. This allows us to simulate a wider range of cluster masses without running into extremely small timesteps which arise from high gas densities (\citealt{courant1928}). We set the gravitational softening of the stars to be $0.001a$ for each simulation. This results in an average softening length of $155$AU throughout our suite. 

We note that the sink particle accretion radius used in H18 $r_{\rm sink}$ is $\approx 1.7$pc. We do not require that all our stars or gas are contained within this radius in these simulations.

The mass of SPH particles in our simulations is $0.06M_\odot$, resulting in a smoothing length $h \approx 0.01$pc initially. Stellar masses are sampled from a Kroupa (\citealt{kroupa2001}) IMF between $0.15M_\odot$ and $100M_\odot$. The gas temperature is chosen to be $10$K, consistent with GMC temperatures from H18. We assign each particle a velocity which we sample using the method outlined in \citet{aarseth}. We calculate the velocity dispersion of our stellar and gas component separately and scale the velocities of each such that both are initially in virial equilibrium ($2K_{s,g}/|P_{s,g}| = \alpha_{s,g} = 1$).

\begin{figure}
    \centering
    \includegraphics[scale=0.35]{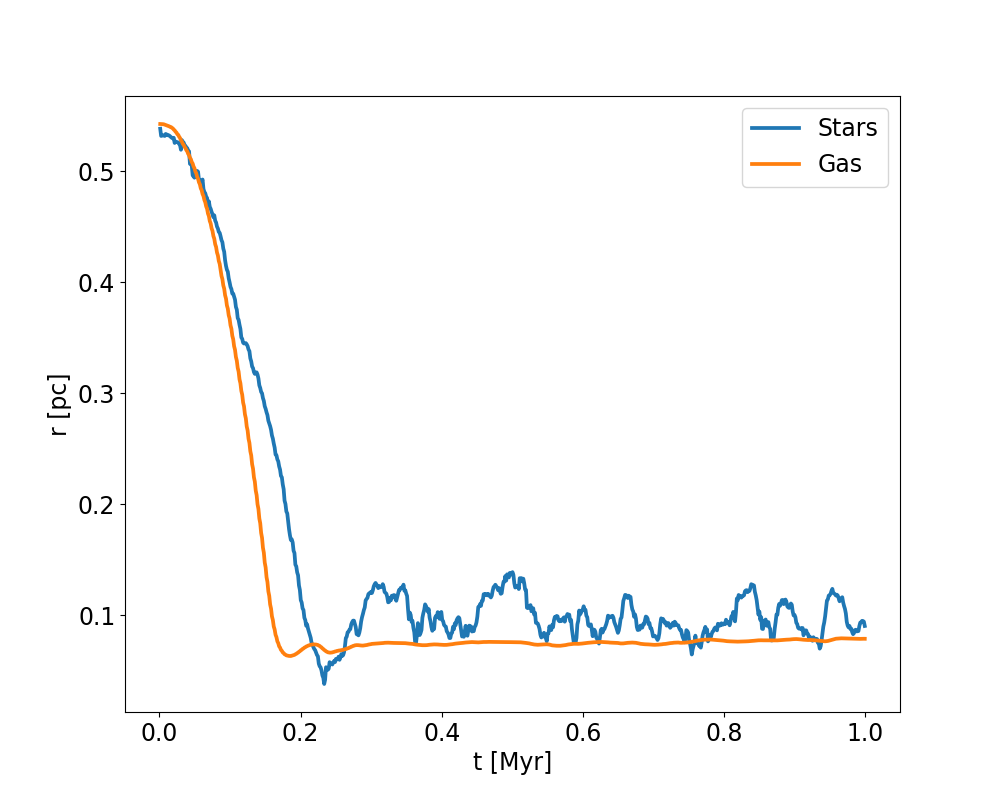}

    \caption{Evolution of the core radii of the stars and gas of an isolated cluster during the numerical relaxation phase.}
    \label{fig:relax}
\end{figure}

\begin{figure}
    \centering
    \includegraphics[scale=0.28]{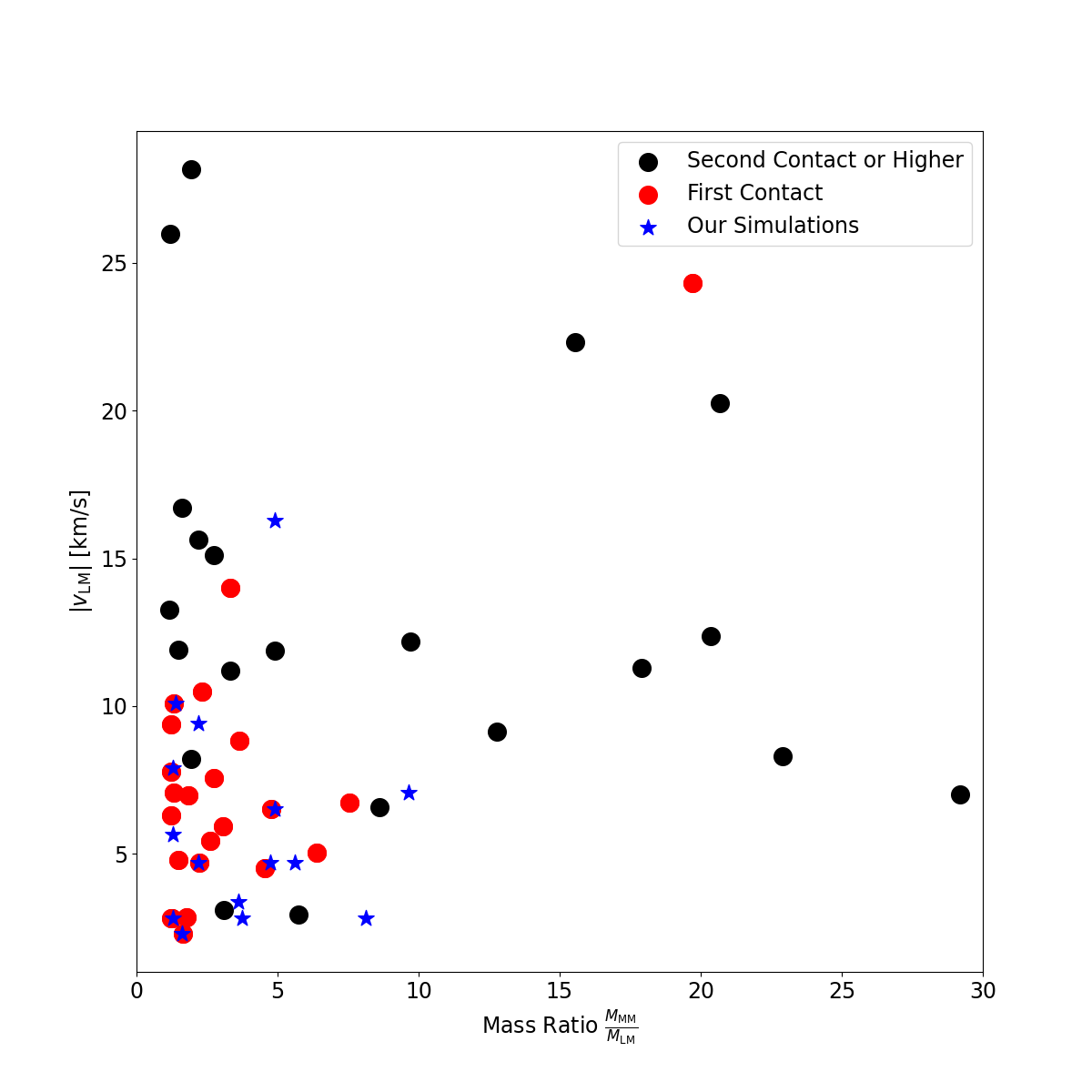}
    \caption{The parameter space from which our collisions are sampled. Circles represent all mergers seen in the H18 simulation, with red points showing all first mergers, and black points showing second and subsequent mergers or higher. Blue stars show the mergers we model for our simulations, some of which are directly taken from H18 (stars on top of red circles).}
    \label{fig:parameter_space}
\end{figure}

Next, we allow our cluster to evolve on its own. We do this so that the stars can have time to react to the new potential introduced by the gas, and vice versa. In order to decide when our clusters are numerically relaxed, we look at the evolution of the core radii for the stars and gas as a function of time, as shown in figure \ref{fig:relax} for a representative cluster. Up until $0.21$Myr, the core radii are decreasing drastically after which both become stable. This effect comes from both stellar and gas components reacting to the potential of the other. In this example we take the snapshot of this cluster after $t = 0.21$Myr as our initial conditions for our merger simulation. This corresponds to roughly 0.01 N-body times. The decrease in the core radius results in $\rho_{hm}$ increasing from the initial $10^3$M$_\odot$pc$^{-3}$. However, the half-mass radius does not change as much as the core radius, and so the value of $\rho_{hm}$ stays below $10^4$M$_\odot$pc$^{-3}$.

\subsection{Merger Setup: The Creation of the Resultant Cluster}
\label{sec:merger}

We select merger pairs from the solar metallicity H18 simulation based on the mass ratio of the pair ($f_M = M_{MM}/M_{LM}$ where $M_{MM}$ and $M_{LM}$ are the total masses of the more and less massive cluster respectively) and the relative velocity of the colliding clusters ($v_{LM}$). A plot of this parameter space from the H18 simulation is shown in figure \ref{fig:parameter_space}. For this paper, we have limited the mass ratio range to be from 0 to 30; there are also 2 mergers that have mass ratios of $\approx 100$ and $\approx 400$. "First contact" mergers, shown as red circles in this figure, are the first merger that occurs for a given star cluster after it has formed. The resultant cluster can then go on to merge with other clusters as the H18 simulation proceeds, and we highlight those mergers as black circles. We simulate 5 of these first contact mergers (labelled run 1-5 in table \ref{tab:collisions} and shown as blue stars on top of red circles in figure \ref{fig:parameter_space}). We choose only first contact mergers for these simulations because we can assume that the star cluster before the merger was a simple system (a Plummer sphere), whereas the resultant cluster after a merger may not be so simple (as we will demonstrate in section \ref{sec:rcp}). These 5 mergers provide us with 10 individual clusters for our initial conditions.

We also simulate mergers that were not present in H18 (shown as isolated blue stars in figure \ref{fig:parameter_space}). To do this, we collide two of our clusters, but not with their counterpart from the H18 simulation. For example, we take the less massive cluster from the run3 pair and simulate its merger with the more massive cluster from the run1 pair in table \ref{tab:collisions} and the run is labelled run3to1. We create other initial conditions by modifying the collision velocity $v_{LM}$ while keeping all other parameters the same as an H18 collision. These runs are labelled as run$x$$\_$$y$v where $x$ is the label of the H18 model run whose velocity we are changing, and $y$ is the factor by which we multiply $v_{LM}$ for that run. 

To set up each collision, we take the relaxed state of the most massive cluster, and place it at $(x,y,z)=(0,0,0)$ with no net velocity. The relaxed state of less massive cluster is placed a distance $x=d$, away from more massive cluster and is given a velocity $v_{LM}$ in the x-direction towards the origin. All of our mergers are head on collisions. We do not include stellar evolution or stellar feedback. Therefore, we should not evolve our simulations past the first expected supernovae as this is the time when most to all of the gas mass has been removed from a star cluster (e.g \citealt{pelupessy2013}, \citealt{chevance}). We run all our simulations for a maximum of $3$Myr.

\begin{table*}
\centering
    \begin{tabular}{c|c|c|c|c|c|c}
         Run number & $M_{MM,s} [10^4M_\odot]$ & $M_{MM,g} [10^4M_\odot]$ & $M_{LM,s} [10^4M_\odot]$ & $M_{LM,g} [10^4M_\odot]$ & $d [pc]$ & $v_{LM} [km s^{-1}]$\\
         \hline
         1 & 0.3 & 2.1 & 0.2 & 1.2 & 2.7 & 2.3\\
         2 & 0.06 & 0.3 & 0.02 & 0.2 & 2.4 & 2.8\\
         3 & 0.1 & 0.8 & 0.06 & 0.4 & 4.9 & 4.7\\
         4 & 0.4 & 1.6 & 0.1 & 0.3 & 2.9 & 6.5\\
         5 & 0.4 & 0.5 & 0.3 & 0.3 & 3.3 & 10\\
         \hline
         2to3 & 0.1 & 0.8 & 0.02 & 0.2 & 2.4 & 2.8\\
         2to5 & 0.4 & 0.5 & 0.02 & 0.2 & 2.4 & 3.4\\
         3to1 & 0.3 & 2.1 & 0.06 & 0.4 & 4.9 & 4.7\\
         3to4 & 0.4 & 1.6 & 0.06 & 0.4 & 4.9 & 4.7\\
         2$\_$2v & 0.06 & 0.3 & 0.02 & 0.2 & 2.4 & 5.6\\
         2to1 & 0.3 & 2.1 & 0.02 & 0.2 & 2.4 & 2.8\\
         2to4 & 0.4 & 1.6 & 0.02 & 0.2 & 2.4 & 7.0\\
         2$\_$2.8v & 0.06 & 0.3 & 0.02 & 0.2 & 2.4 & 7.8\\
         3$\_$2v & 0.1 & 0.8 & 0.06 & 0.4 & 4.9 & 9.4\\
         4$\_$2.5v & 0.4 & 1.6 & 0.1 & 0.3 & 2.9 & 16
    \end{tabular}
    \caption{Parameters for our H18 merger runs. Column 1: the name of the run, column 2: the mass of the more massive cluster in stars, column 3: mass of the more massive cluster in gas, column 4: mass of the less massive cluster in stars, column 5: mass of the less massive cluster in gas, column 6: the initial separation along the $x$-axis of the two clusters, column 7: the x-velocity kick given to the less massive cluster towards the origin. The horizontal line splits the models taken directly from H18 sinks and those given some alteration. Runs are ordered by increasing collisional velocity.}
    \label{tab:collisions}
\end{table*}

\begin{figure*}
    \centering
    \includegraphics[scale=0.24]{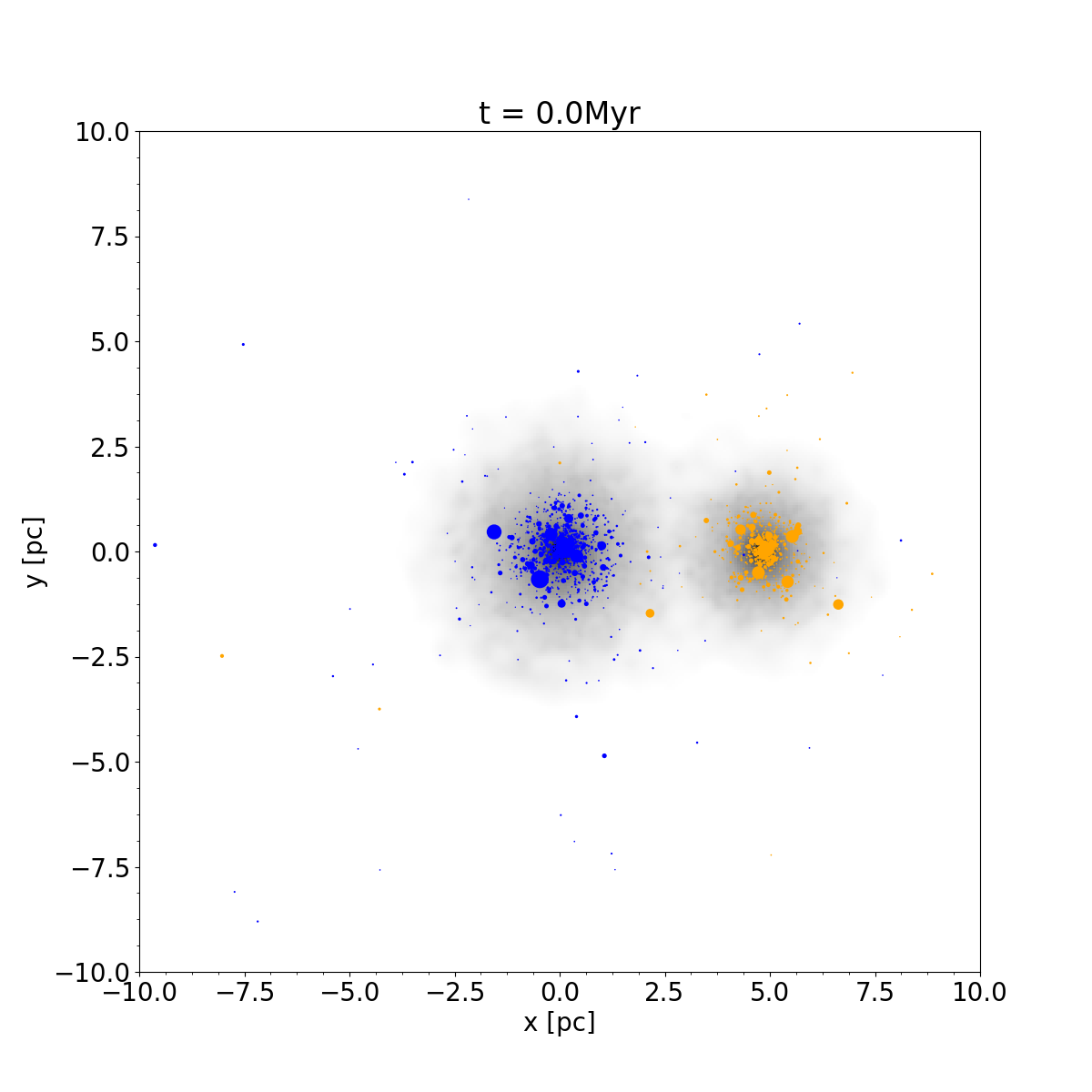}
    \includegraphics[scale=0.24]{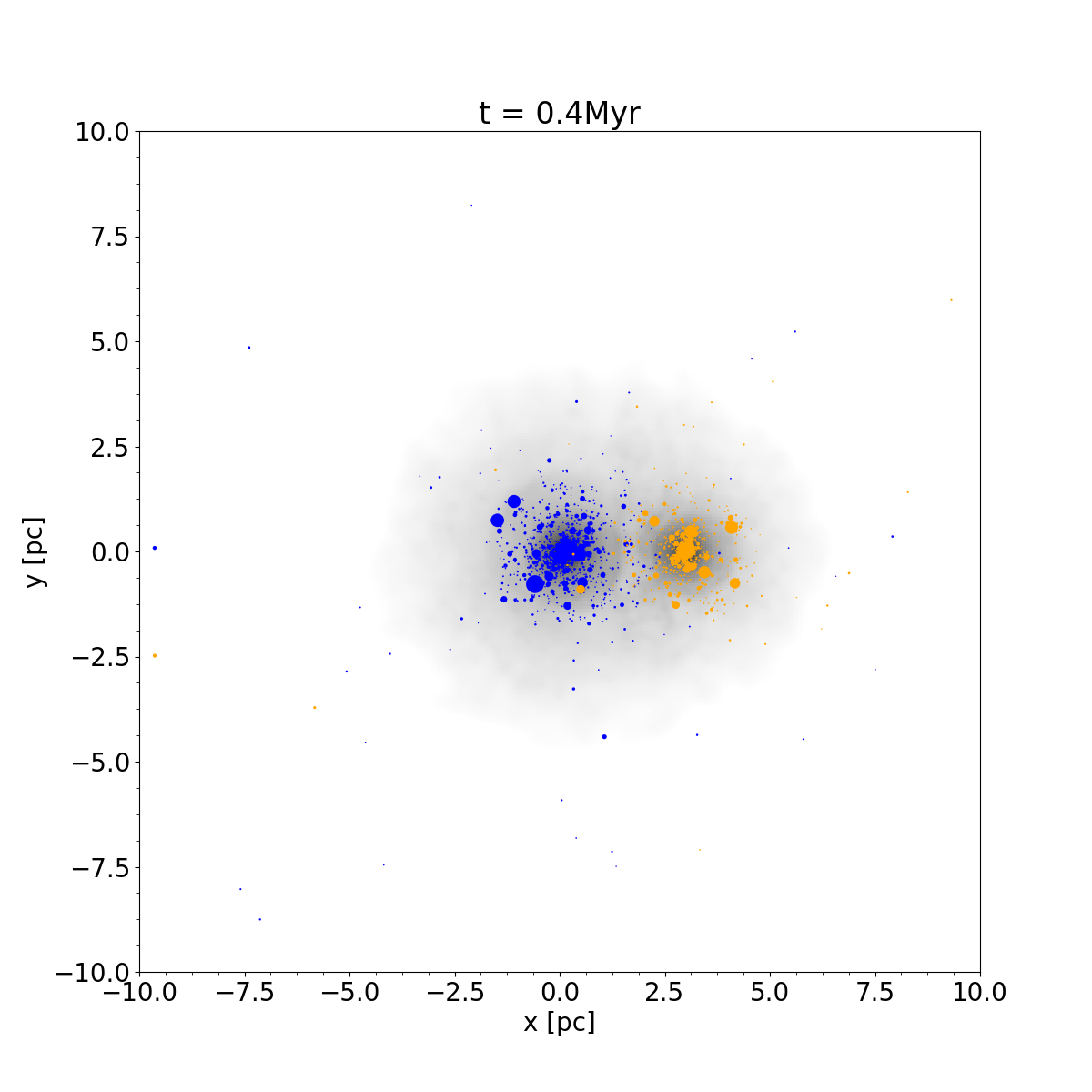}
    \includegraphics[scale=0.24]{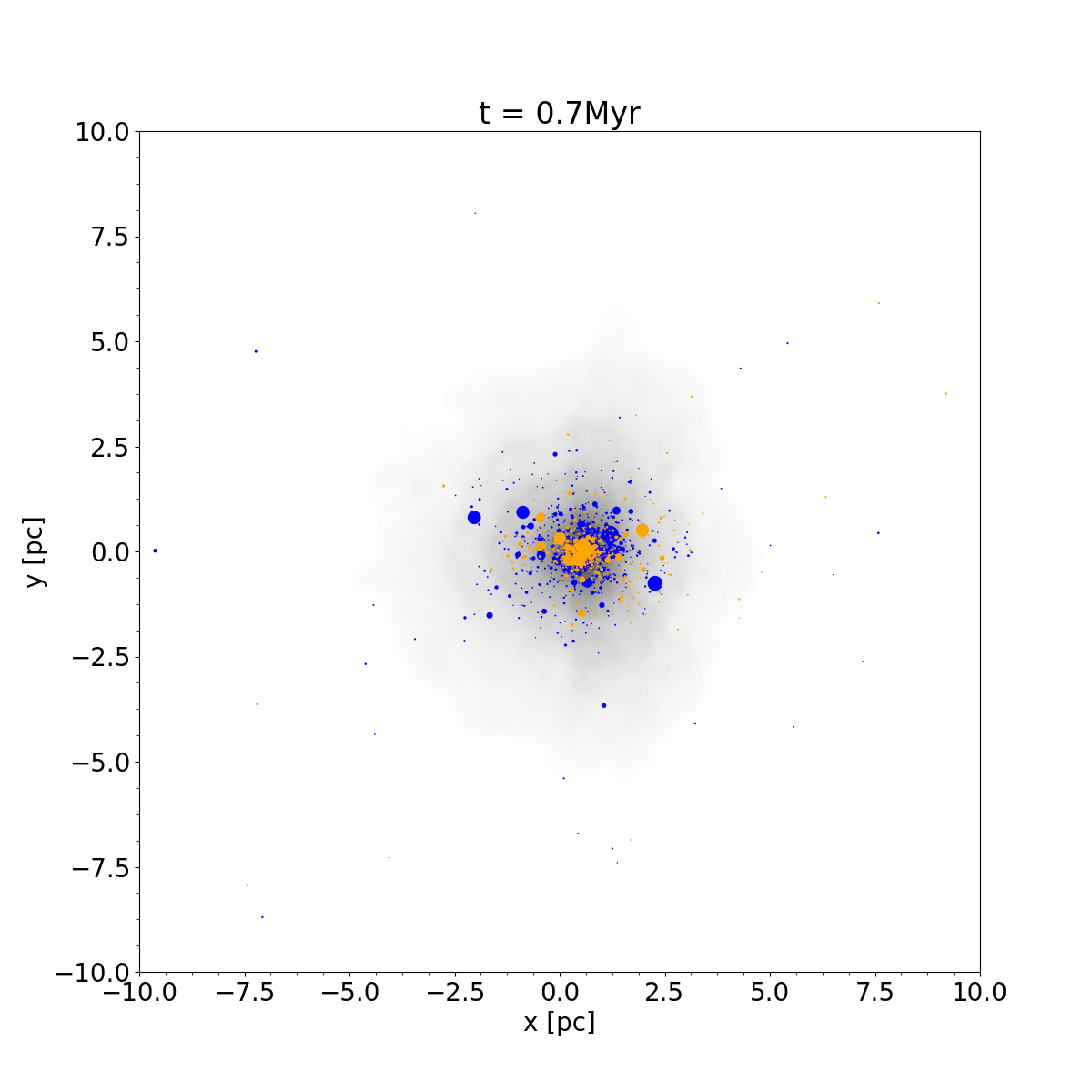}
    \includegraphics[scale=0.24]{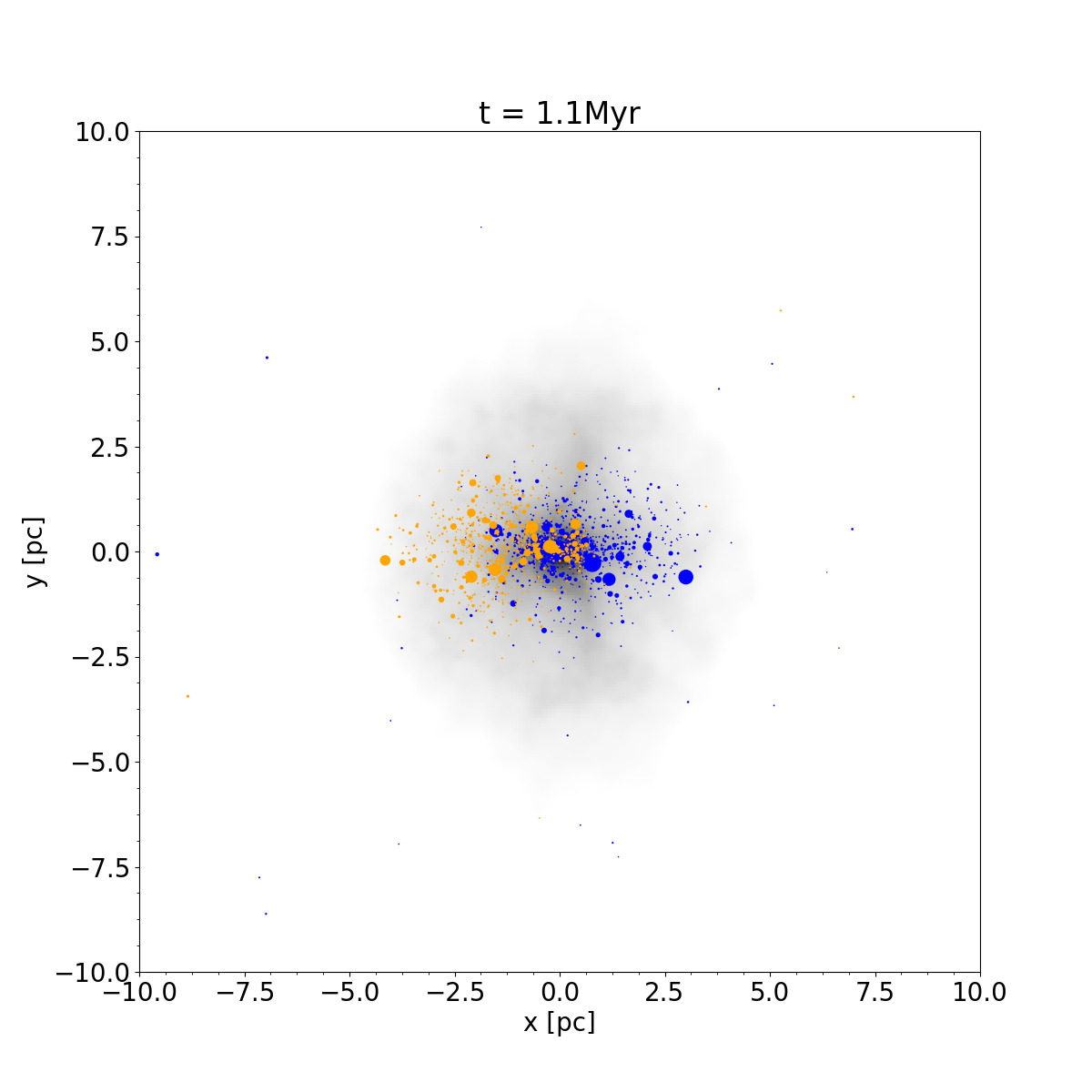}
    \includegraphics[scale=0.24]{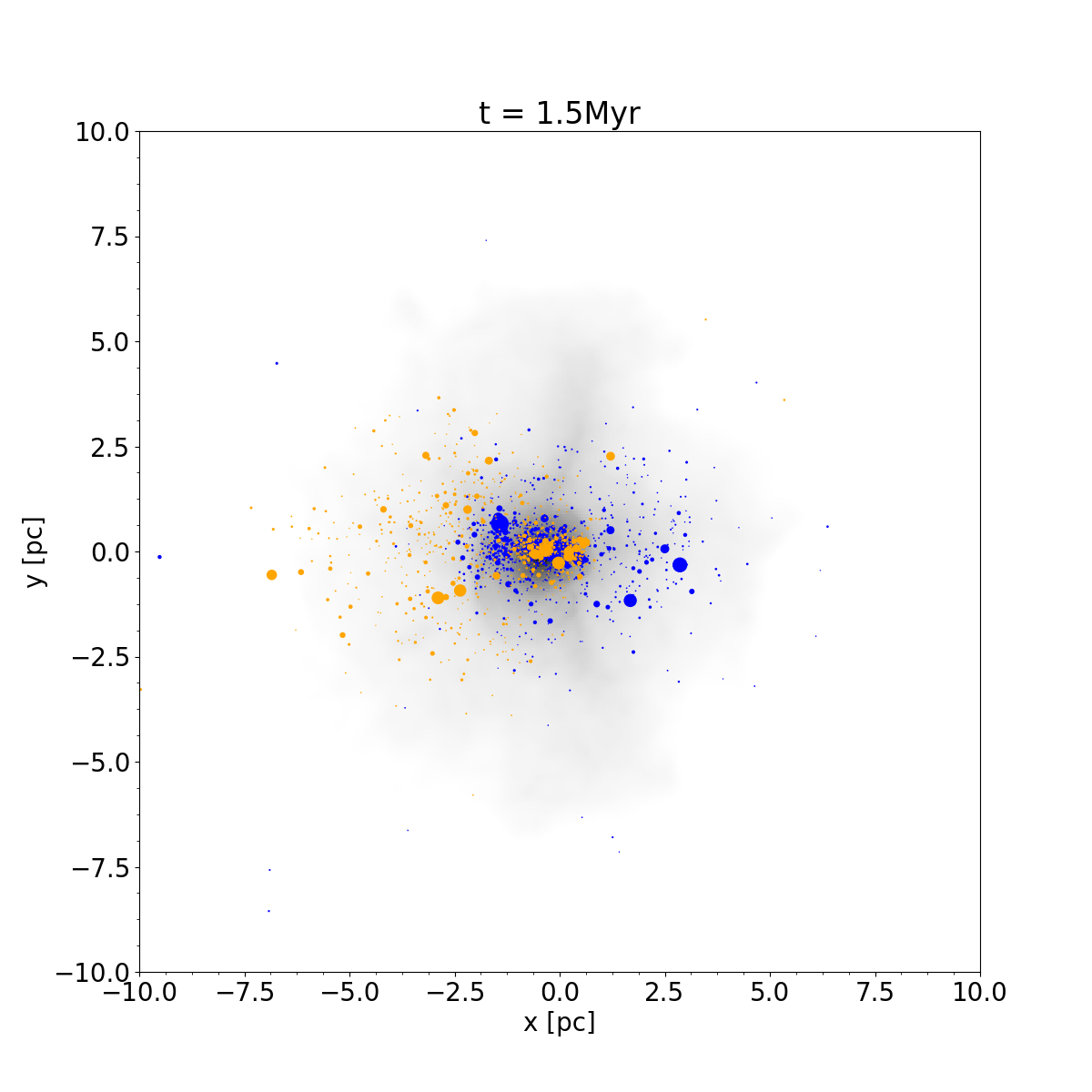}
    \includegraphics[scale=0.24]{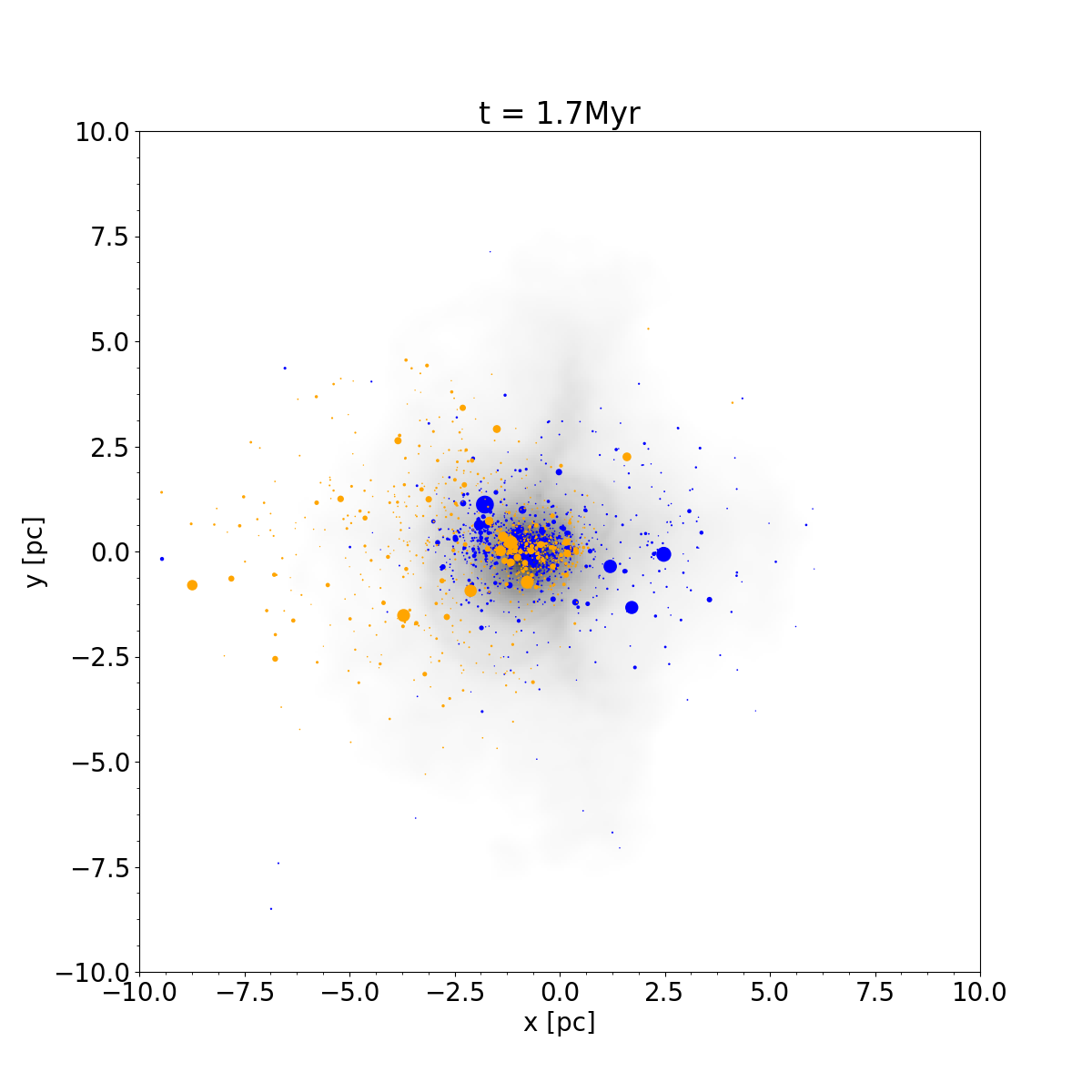}
    \caption{Snapshots of the stars and gas from merger run3 leading up to the monolithic time. The circles represent the stars from each cluster (orange belong to less massive cluster, blue belong to more massive cluster) and the gas is shown in black with darker regions showing gas with higher density. Size of the filled circles scales with mass of the star. Time is taken with respect to the beginning of the simulation. The left panel on the second row shows the merger at $t = 0.7$Myr $= t_{col}$ (the collision time) and the right panel on the second row shows the merger at $t = 1.1$Myr $= t_{mon}$ (the monolithic time).}
    \label{fig:model1}
\end{figure*}

\section{Resultant Cluster Properties}
\label{sec:rcp}
We begin with a discussion of a typical merger from our sample (run3 in table \ref{tab:collisions}). The total masses of the more and less massive clusters in the merger are 0.9 and 0.5 $\times 10^4$M$_\odot$, mostly consisting of gas for this particular run. This mass ratio of $f_{\rm M} \approx 2$ and a collisional velocity of $v_{LM} = 4.7$kms$^{-1}$ places this merger in the region of parameter space where we find most of the first contact mergers from H18. Most of the mass of both clusters resides within $r_{sink} \approx 1.7$pc (80\% of the more massive cluster mass, 90\% of the less massive cluster's mass). Snapshots of this merger can be seen in figure \ref{fig:model1}. The gas density is shown as the greyscale with a minimum at 1 M$_\odot$pc$^{-3}$ and a maximum of 10$^5$M$_\odot$pc$^{-3}$, while the stars are shown as blue or orange circles.

At $t \approx 0.7$Myr, the centres of the gas distributions for the two clusters overlap for the first time. We denote this time as the collision time $t_{col}$. From $t_{col}$ until the end of our simulation, the gas remains as a single entity. However, the stars behave more like a collisionless system, and the lower-mass cluster stars pass through the higher-mass cluster before either being dispersed or returning towards the centre of mass. At about $0.4$Myr after the collision a single monolithic cluster is formed. We denote this time as $t_{mon}$, and calculate it by determining the centre of the inner 5\% of stars (approximately the percentage of stars within the core radius of the original clusters). When the difference between those two centres remains smaller than the original core radius of the most massive cluster, we say that the resultant cluster is monolithic.

We can now analyze properties of the stellar and gas components of the resultant cluster after the merger has taken place.

\subsection{Bound Fraction}
\label{sec:size}

\begin{figure}
    \centering
    \includegraphics[scale=0.3]{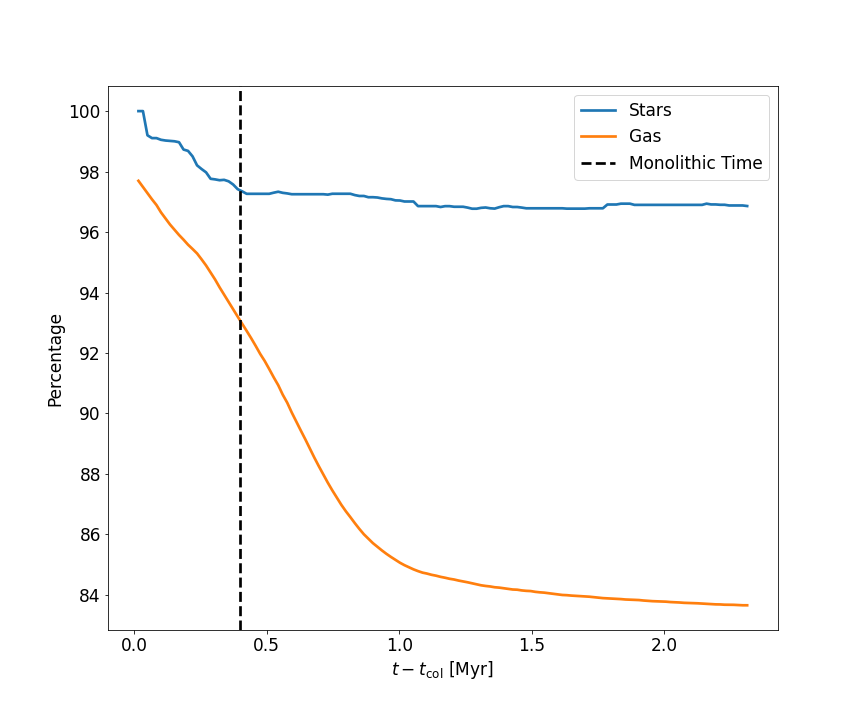}

    \caption{Bound mass percentages of the resultant cluster's stellar (blue) and gas (orange) components after the collision. The black dashed line shows the monolithic time of this simulation.}
    \label{fig:unbound}
\end{figure}

\begin{figure*}
    \centering
    \includegraphics[scale=0.35]{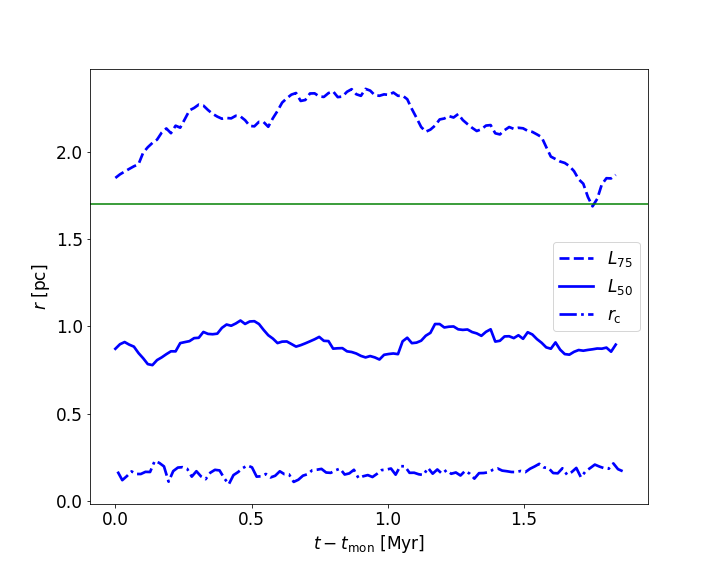}
    \includegraphics[scale=0.35]{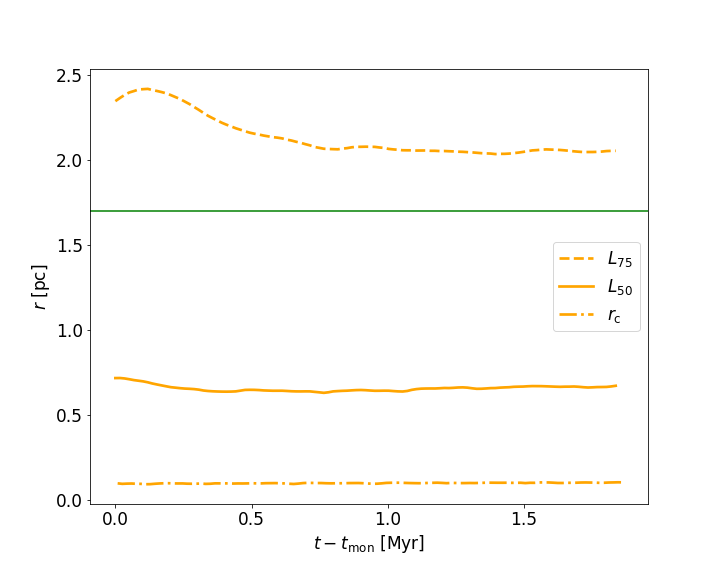}

    \caption{$75$\% mass radii (dashed), half mass radii (solid), and core radii (dot-dashed) of bound stellar (left) and gas (right) component for the resultant cluster after the cluster has become monolithic. The green line indicates the sink radius prescribed in the H18 simulations ($r_{sink} \approx 1.7$pc).}
    \label{fig:bounds}
\end{figure*}

A property of the sink particle prescription used in H18 is that all mass inside each sink remains in the sink throughout the entire simulation. When two parent sink particles merge, the resultant sink mass will simply be the sum of the masses of each parent. We can test this approximation by looking at the bound percentage of stellar and gas mass in the resultant cluster after the merger takes place. This can be seen in figure \ref{fig:unbound} for run3. We consider a particle bound when its $T + U < 0$ where $T$ is its kinetic energy, and $U$ is the potential of both stars and gas. We see that by the end of this simulation, $\approx 3$\% of stellar mass and $\approx 16$\% of gas mass has become unbound. Therefore, we conclude that for this collision, the assumption of no mass loss in the sink particle prescription is reasonable. Most of the decrease in bound stellar mass happens between $t_{col}$ and $t_{mon}$, as stars that initially belonged to the less massive cluster are flung out of the gravitational pull of the resultant cluster. These results are consistent with simulations done by \citet{BK2015} who see a negligible amount of unbound stellar mass in merger simulations with an analytic gas potential. As well, N-body simulations done by \citet{tvr2} show that $\lessapprox$ 10\% of their stellar mass becomes unbound after the merger of two equal mass Plummer spheres.

\subsection{Cluster Size}
\label{sec:cluster_size}
We can now look at the Lagrangian radii of the resultant cluster. We look at the core, half mass, and 75\% mass radii of the bound mass (stars or gas) in the resultant cluster, and compare them to the H18 sink radius of 1.7 pc in figure \ref{fig:bounds}. We see that the core radius remains stable for the entire simulation after $t_{mon}$ at $r_c \approx 0.2$pc and $r_c \approx 0.1$pc for the stellar and gas components respectively. Similarly, the half-mass radii remain approximately constant as well, and are well below the sink radius of 1.7 pc. However, if we look at the 75\% Lagrangian radius of each component, we see that they are both greater than $r_{sink}$. Therefore, a significant amount of the resultant cluster mass is outside the boundary set in the H18 simulation.

\begin{figure}
    \centering
    \includegraphics[scale=0.35]{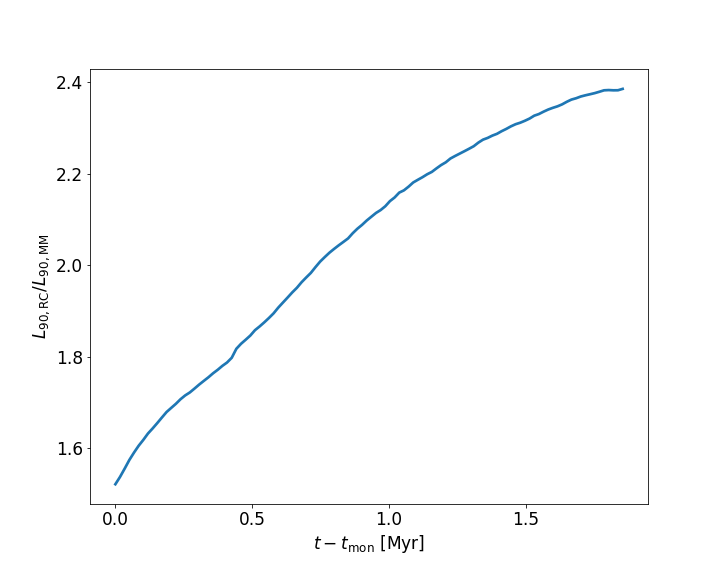}
    \caption{Time evolution of the 90\% Lagrangian radius for the total (stars + gas) resultant cluster, normalized to that same radius for the most massive parent cluster.}
    \label{fig:new_sink}
\end{figure}

To make the comparison with the H18 simulation a little clearer, we show the evolution of the 90\% Lagrangian radius of the entire resultant cluster (bound stars and gas together), normalized by the 90\% Lagrangian radius of the more massive cluster of the collision, in figure \ref{fig:new_sink}. The outermost region of the cluster continues to grow to the end of our simulation. Because the sink particle prescription allows gas accretion to take place if gas is within one $r_{sink}$ and gravitationally bound to the cluster, we conclude that a sink particle prescription which does not allow this growth in radius may be accreting less gas than it should, and may result in a smaller final mass for the clusters formed in H18. The magnitude of this discrepancy is dependant on how prompt the secondary merger process is. If the resultant cluster merges with another cluster quickly after it is formed, the cluster will not have had enough time to significantly grow, meaning that only a small fraction of extra gas would be added to the resultant sink. However, the longer it takes for a resultant cluster to undergo another merger, the larger the cluster grows and the more mass it may accrete.

\subsection{Best Fit Radial Density Profiles}
\label{sec:dens_prof}
We now turn our attention to the radial density profiles of the resultant cluster. In this analysis we treat the stars and gas of the resultant cluster separately and include only the bound members of each component. It would be convenient for future simulations if we could identify an analytic or simple density profile for the result of a cluster merger. Therefore, we fit commonly used density profiles to the bound stellar and gas components of the resultant cluster. We consider the Plummer, \citet{eff} (EFF) model, and the \citet{king} models for our fits. Plummer models have been used in a variety of numerical simulations to initialize stars and gas (e.g \citealt{pelupessy2013}, \citealt{BK2015}, \citealt{sills2018}) as have King models (e.g \citealt{gb2001}, \citealt{king_ex1}). The EFF model is given by
\begin{equation}
    \rho(r) = \rho_{0}(1 + \frac{r^2}{a^2})^{-{\frac{\gamma+1}{2}}}
    \label{eq:eff}
\end{equation}
The two parameters allowed to vary above are the scale radius $a$, and the slope of the density tail $\gamma$. Note that the Plummer density profile is a special case of \ref{eq:eff} with $\gamma = 4$.  The \citet{king} model has two parameters that are allowed to vary: the tidal radius of the cluster $r_t$ and the dimensionless central potential $W_0$. We calculated the King density profile using \texttt{galpy}\footnote{\href{https://github.com/jobovy/galpy}{https://github.com/jobovy/galpy}} (\citealt{galpy}) and to calculate the density profile of our simulated resultant cluster, we use the \texttt{python} package \texttt{Clustertools}\footnote{\href{https://github.com/webbjj/clustertools}{https://github.com/webbjj/clustertools}} which creates radial bins of equal numbers of particles and calculates the density in each bin. Our bins are centered on the density centre (\citealt{zwart})). For both the stellar and gas components, we find that our choice of density centre over centre of mass does not greatly affect our best fit parameters for any of our models.

\begin{figure}
    \centering
    \includegraphics[scale=0.35]{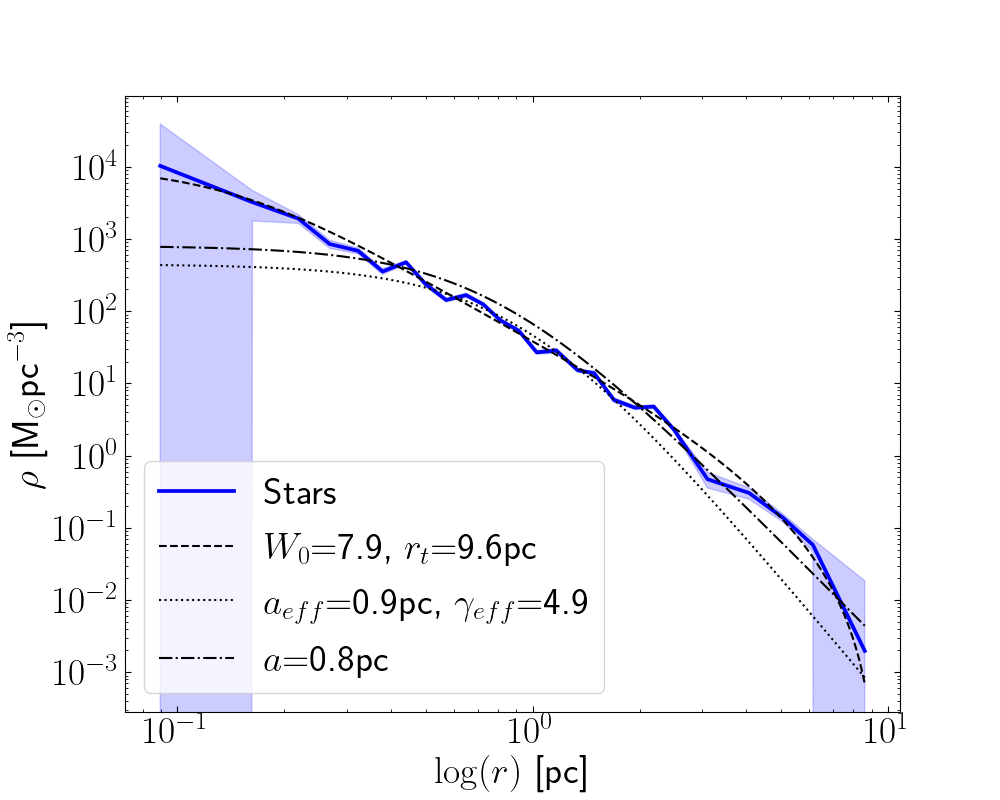}
    \includegraphics[scale=0.35]{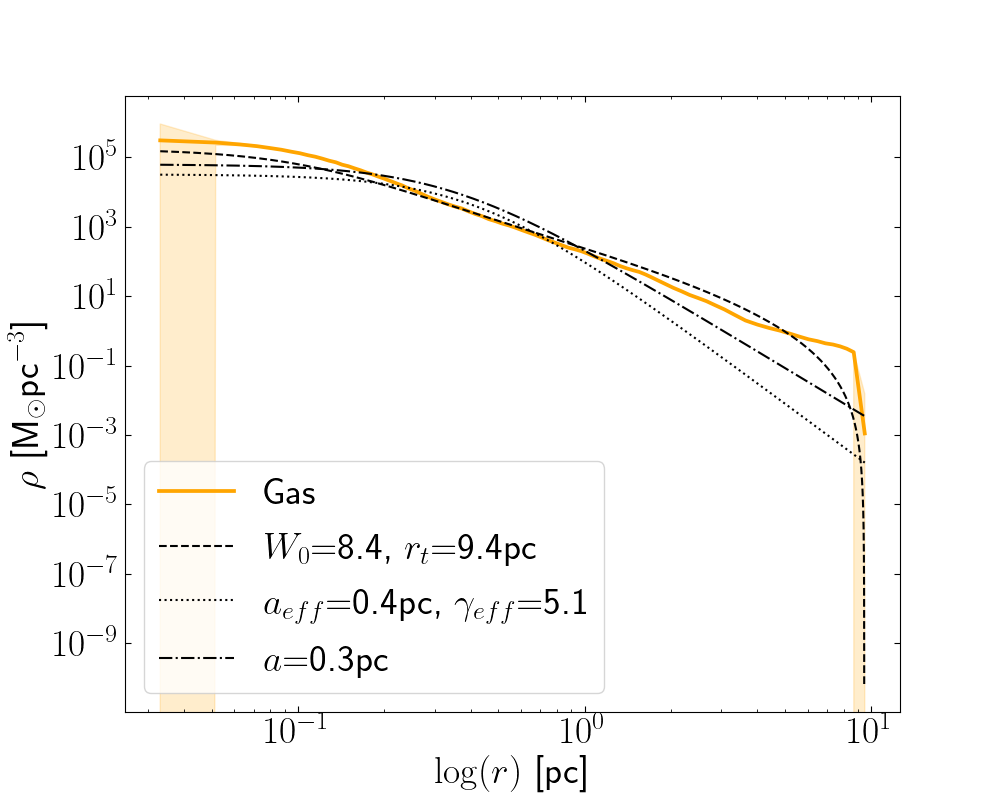}
    \caption{Stellar and gas radial density profiles plotted with best fit profiles from three theoretical functions: Plummer (dot-dashed), EFF (dotted), and King (dashed). The shaded regions show one standard deviation in the density calculation for a given radial bin.}
    \label{fig:dens_sample}
\end{figure}

We choose radial bins containing 80 stars or 2000 gas particles. We then use the \texttt{python} function \texttt{curve\_fit} from \texttt{scipy} (\citealt{scipy}) which uses the Levenberg-Marquardt (LM) method (\citealt{LM}) to calculate the best fit parameters for each model. We performed this fit from $t_{mon}$ to the end of the simulation. We show a snapshot of the resultant cluster's density profile overplotted with the three best fit functions at one timestep in figure \ref{fig:dens_sample}. We find that the general shape of the profiles (high density in the centre, low density in the outskirts) matches all of the models.

Compared to our initial Plummer profile, we find that the merger process has steepened the surface density in the outskirts of our stellar and gas density profiles (a $\gamma = 4$ profile becomes closer to a $\gamma = 5$ profile). In the inner regions, both the stars and the gas have smaller cores. The $W_0 = 8$ for the stellar King profile and $W_0 = 9$ for the gas King profile are consistent with a more centrally concentrated cluster.

However, none of the models ever provide a satisfactory fit to the simulated density profile. We conclude that these three analytical models cannot be used to describe our resultant cluster after the merger process. In order to achieve a good representation of such a cluster, we must continue to use the results of our numerical simulations directly.

\subsection{Star Formation Density Thresholds}

\begin{figure}
    \centering
    \includegraphics[scale=0.35]{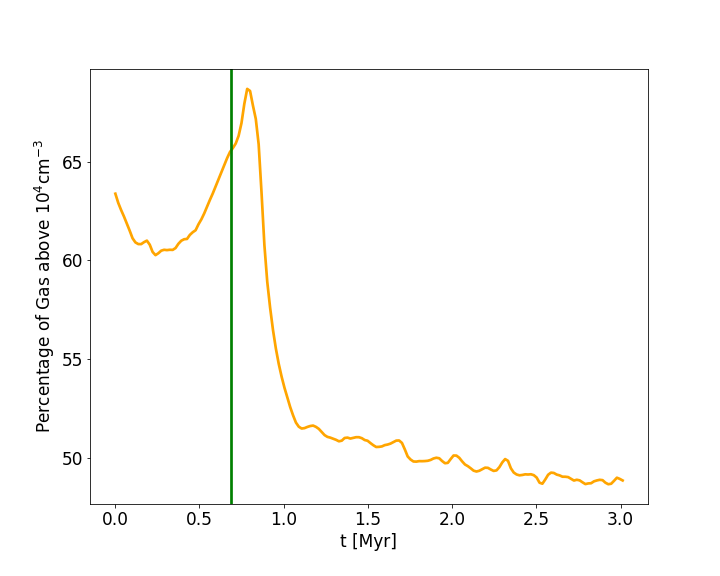}

    \caption{Percentage of gas above $10^4$cm$^{-3}$ throughout run3. The green line shows the collision time $t_{col}$ for this merger simulation.}
    \label{fig:sf_densn}
\end{figure}

In the H18 simulations, the sinks form in gas above a density threshold of 10$^4$ cm$^{-3}$ and it is assumed that all the gas inside the sink has that density for the rest of the simulation. Star formation in H18 takes place with a constant efficiency per freefall time assuming that density. Their sinks are sampled every 0.36 Myr to determine how many stars should form. We track the amount of dense gas in our clusters and how that changes throughout the run3 merger. In particular, we look at the gas with densities above 10$^{4}$cm$^{-3}$ and 10$^{5}$cm$^{-3}$ which are commonly quoted density ranges above which dense star-forming cores begin to form (e.g \citealt{evans09}, \citealt{heid10}, \citealt{lada2010}, \citealt{lada2012}). We show the percentage of gas above 10$^4$cm$^{-3}$ in figure \ref{fig:sf_densn}.

At the beginning of our simulation, the amount of gas above 10$^4$cm$^{-3}$ is only about 65\%, compared to the 100\% assumed for the H18 sink particles. This percentage increases just after the moment of collision (green vertical line), but does not last for long compared to the timescale at which H18 samples their sinks to form stars (0.36Myr). Shortly after the collision time, only about half of the gas in the cluster has density above 10$^4$cm$^{-3}$. Therefore we see that for run3, the merger process results in a decrease in potentially star forming gas in the resultant cluster. Gas above 10$^5$cm$^{-3}$ shows a very similar trend. This trend of an increase in star forming gas followed by a decline in such gas was also found in the star cluster formation simulations presented in \citet{fujii2021}.

\section{Exploring the Parameter Space}
\label{sec:suite}
\begin{figure}
    \centering
    \includegraphics[scale=0.5]{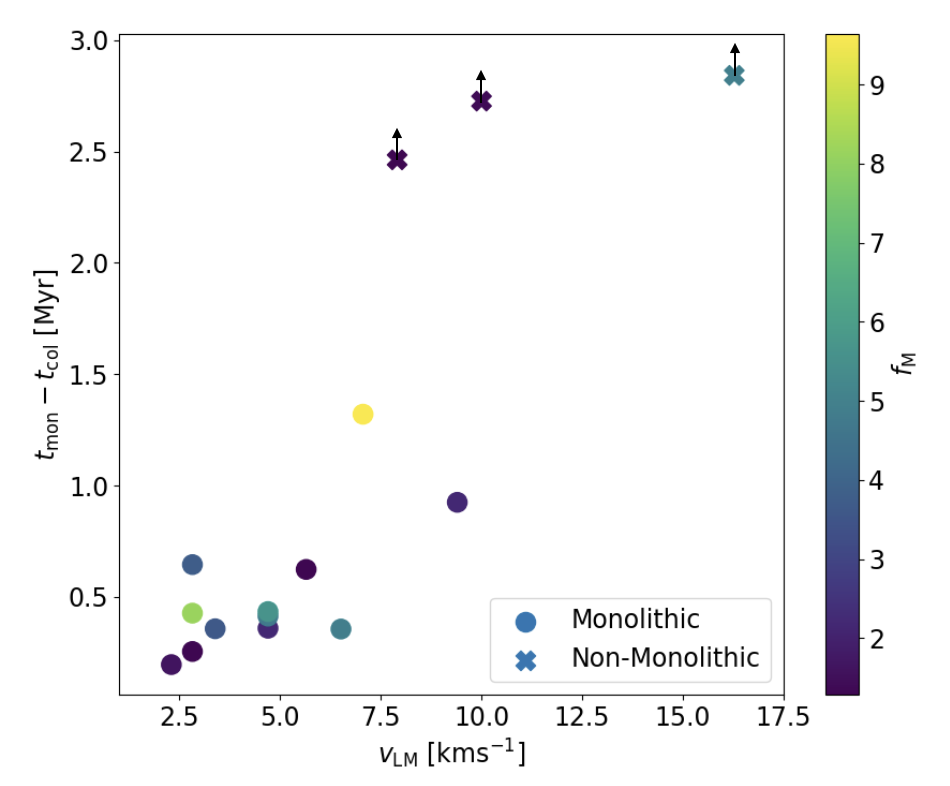}
    \caption{Time after the collision for the two clsuters to become one monolithic structure, plotted against the collisional velocity of each simulation. The colour bar gives the mass ratio of each merger. We show both the monolithic ($t_{mon} < t_{sim}$) simulations as solid circles and the non-monolithic ($t_{mon} > t_{sim}$) simulations as crosses with arrows. The arrows indicate that this is a lower limit of the monolithic time for the non-monolithic simulations.}
    \label{fig:t_mon_par}
\end{figure}

In this section, we present the results of our full suite of simulations. We first show the time needed to form a single monolithic cluster as a function of the collisional velocity in figure  \ref{fig:t_mon_par}.  We note that three simulations never form a single cluster ($t_{mon}$ exceeds the length of the simulation $t_{sim}$), indicated on the plot with "x" symbols and arrows. We will discuss those in section \ref{sec:non-monolithic}.

We find that, as expected, a higher collisional velocity leads to longer time for the two clusters to fully merge. The relationship between $t_{mon} - t_{col}$ and the mass ratio ($f_M$) is less clear, but in general a higher mass ratio results in a longer time for the resultant cluster to become monolithic. We note that in the H18 simulations, the average time between first and subsequent collisions is about 0.4 Myr, and therefore many of our mergers simulated here will not yet have resulted in monolithic clusters before their next encounter occurs. This means it is even more critical that we use detailed simulations of cluster mergers to understand the build-up of massive clusters.

\subsection{Monolithic Resultant Clusters}

We find that for most of our mergers, $\lesssim 5$\% of the stellar mass becomes unbound from our resultant cluster, as long as the collision creates a single resultant cluster. Mergers with higher collisional velocity and lower mass ratios tend to lose more mass. In our most extreme monolithic simulation, run3\_2v, which has a collisional velocity of $v_{LM} \approx 9.4$kms$^{-1}$ and mass ratio $f_M \approx 2$, we lose $12$\% of the stellar mass. The unbound gas mass percentage behaves in a similar way (strong dependence on $v_{LM}$ and no clear dependence on $f_M$). However, the unbound gas mass percentage never exceeds 20\% of the initial gas mass.

\begin{figure*}
    \centering
    \includegraphics[scale=0.45]{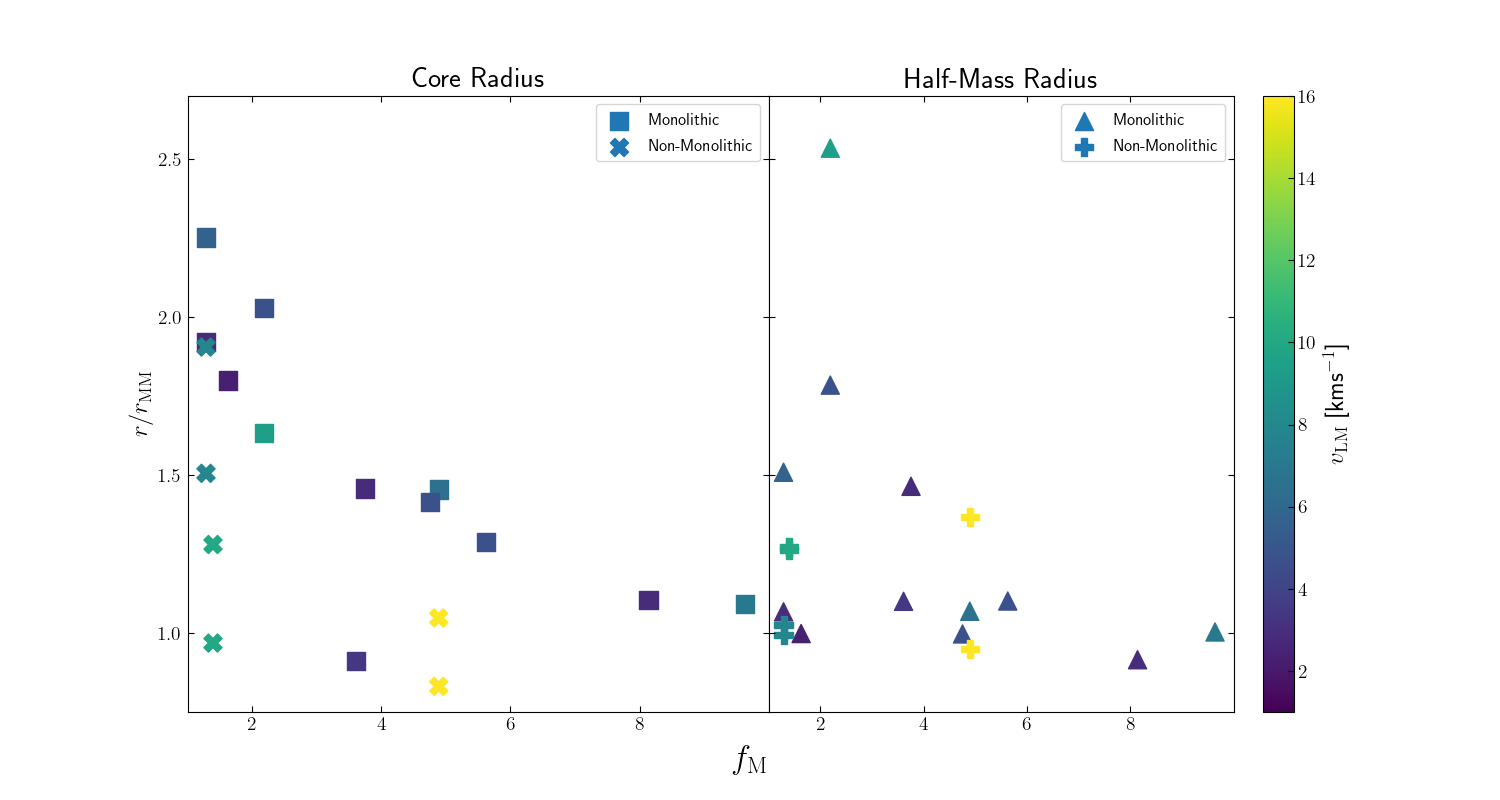}
    \caption{Inner Lagrangian radii of the stellar component of the resultant cluster, normalized by that radius of the more massive parent cluster, plotted against mass ratio for monolithic and non-monolithic simulations. The colour bar shows the collisional velocity of given merger.}
    \label{fig:inner}
\end{figure*}

We show the core radii and half-mass radii of the stars at 1 Myr after the monolithic time, as a function of the mass ratio of the merger and the collisional velocity, in figure \ref{fig:inner}. We find that most of the resultant clusters are larger than their original parent cluster, and there is a strong dependence of this growth on the mass ratio. At low $f_M$, because the masses of both clusters are similar, the core of the resultant cluster has gained a significant amount of new stellar mass. As we get to higher $f_M$ values, we are looking at more minor mergers and the resultant cluster's growth is less affected by the less massive cluster, causing the plateau at $r/r_{MM} \approx 1$ for high $f_M$. We find that the collisional velocity does not have an effect on the growth of the core radius, but does have a small effect on the half mass radius.

The gas component of the resultant clusters is also larger than that of the parent clusters, but in this case we find that the growth of the gas component is sensitive to both the collisional velocity and the mass ratio.

\begin{figure}
    \centering
    \includegraphics[scale=0.35]{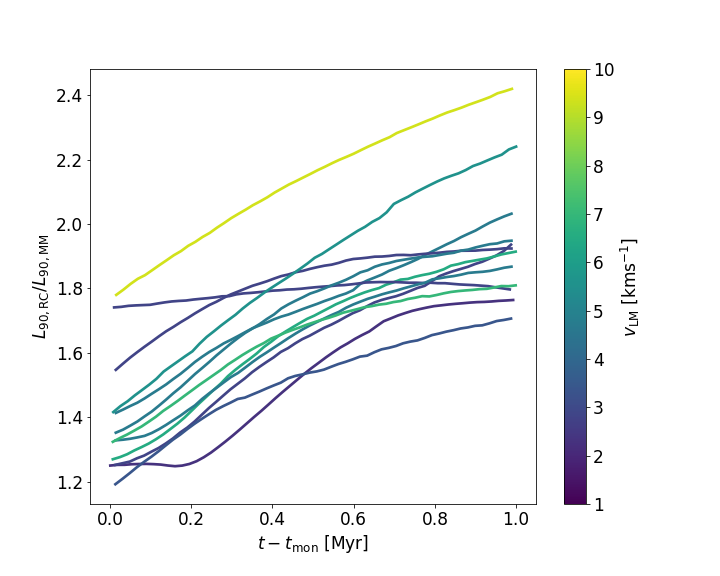}
    \caption{The 90\% Lagrangian radii of the entire bound stellar and gas content of the resultant cluster normalized by that of the entire more massive cluster as a function of time after $t_{mon}$ for every merger whose resultant cluster is one monolithic structure. The colour bar shows the collisional velocity of the merger.}
    \label{fig:L90}
\end{figure}

We show the values of the 90\% Lagrangian radius for the full resultant cluster (stars and gas) for each simulation, normalized by that simulation's $L_{90,MM}$ plotted for all monolithic simulations in figure \ref{fig:L90}. The colour bar shows the collisional velocity of the merger. Most of our monolithic clusters continue to expand for the duration of our simulation, suggesting that a constant sink radius may not be an accurate description of merged clusters.

We find that the density profiles of all our monolithic clusters have similar shapes as those shown for run3 in figure \ref{fig:dens_sample} and that all stellar components show a high central concentration (King profile $W_0 \approx 8$). However, as expected, none of the simple density profiles provide good fits to either the stellar or gas density profile. We conclude that any simulations which follow the building of a massive cluster from the merger of subclusters should use computed profiles of merged clusters at the appropriate time, rather than being able to use simpler models for the results of the first merger.

\subsection{Non-Monolithic Simulations}
\label{sec:non-monolithic}
\begin{figure*}
    \centering
    \includegraphics[scale=0.24]{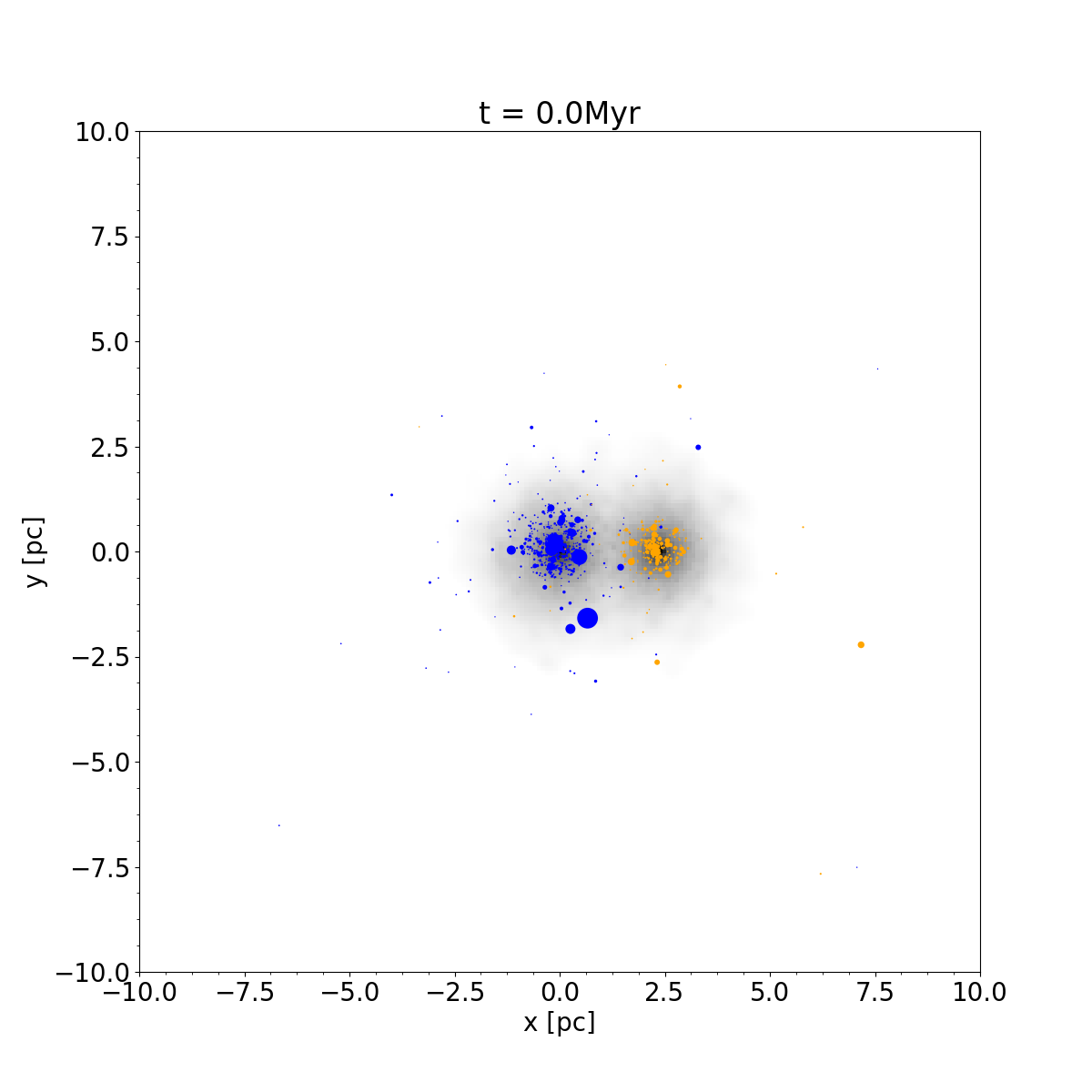}
    \includegraphics[scale=0.24]{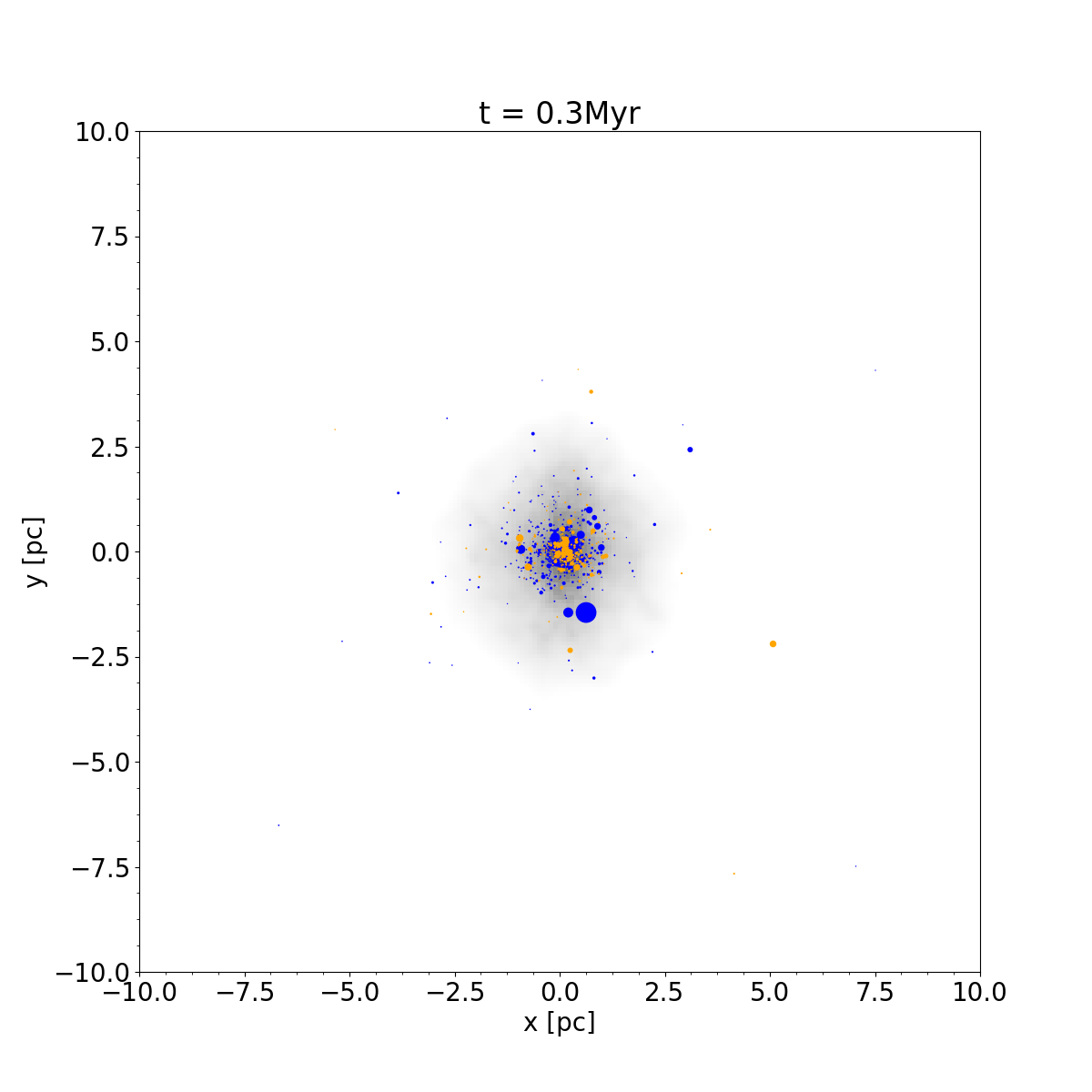}
    \includegraphics[scale=0.24]{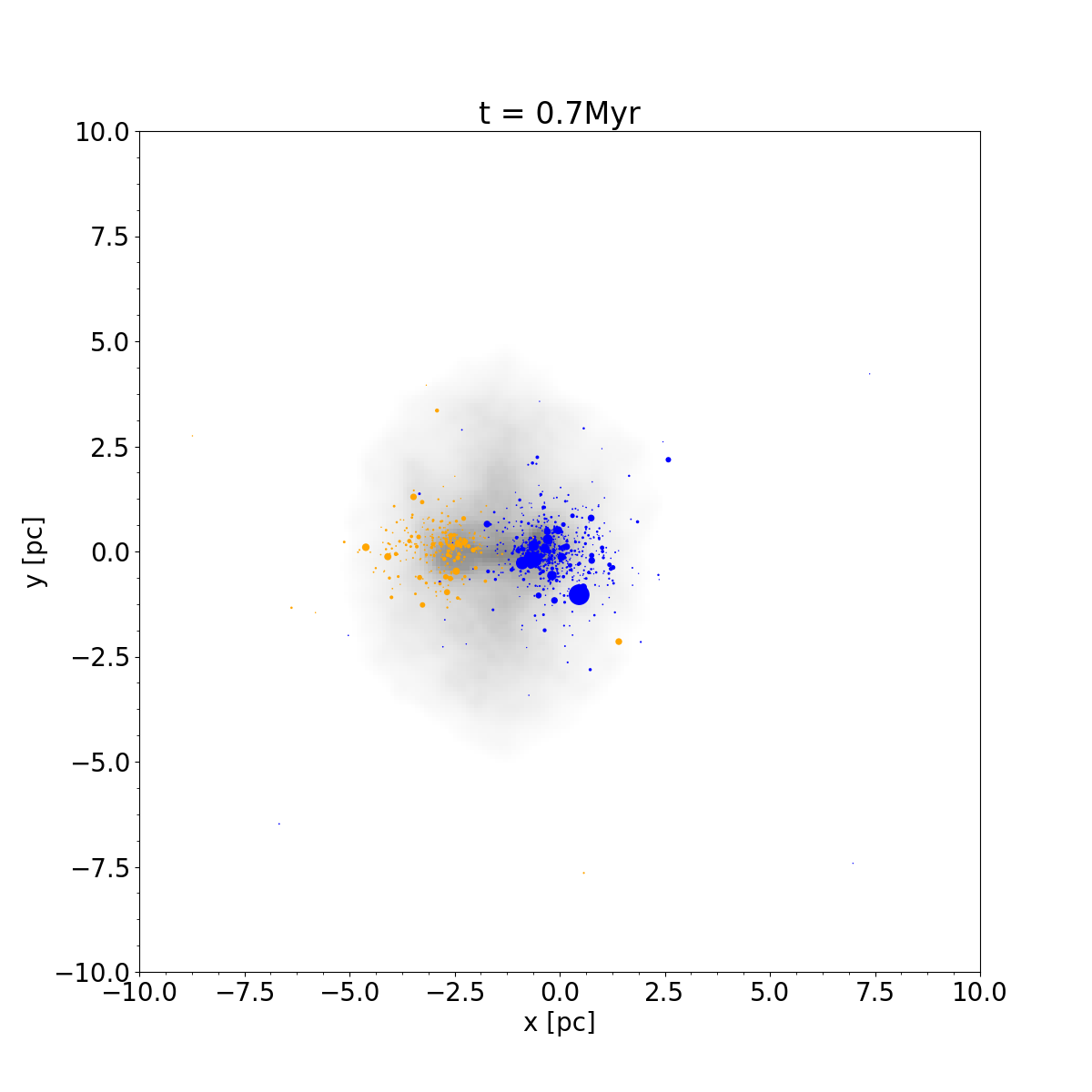}
    \includegraphics[scale=0.24]{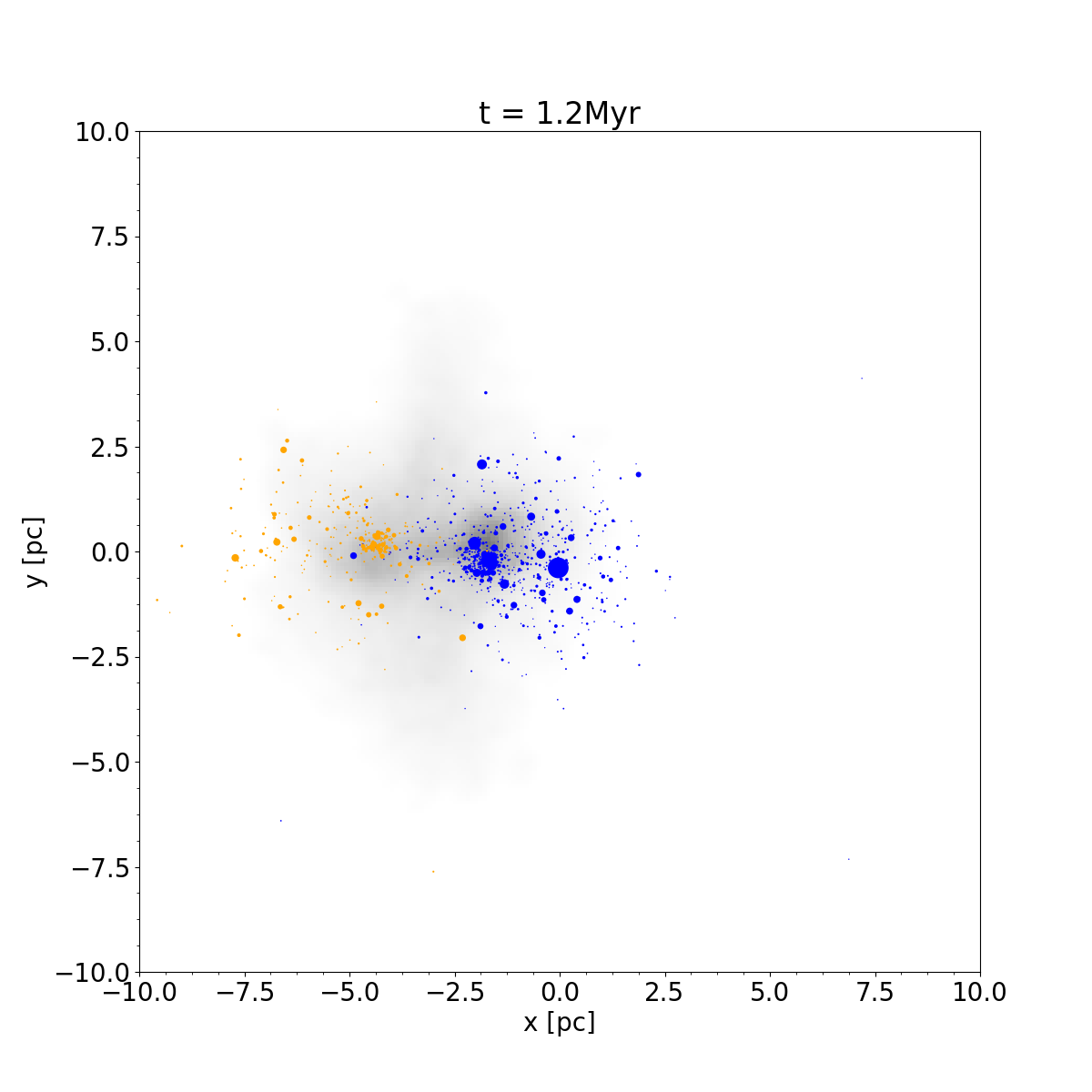}
    \includegraphics[scale=0.24]{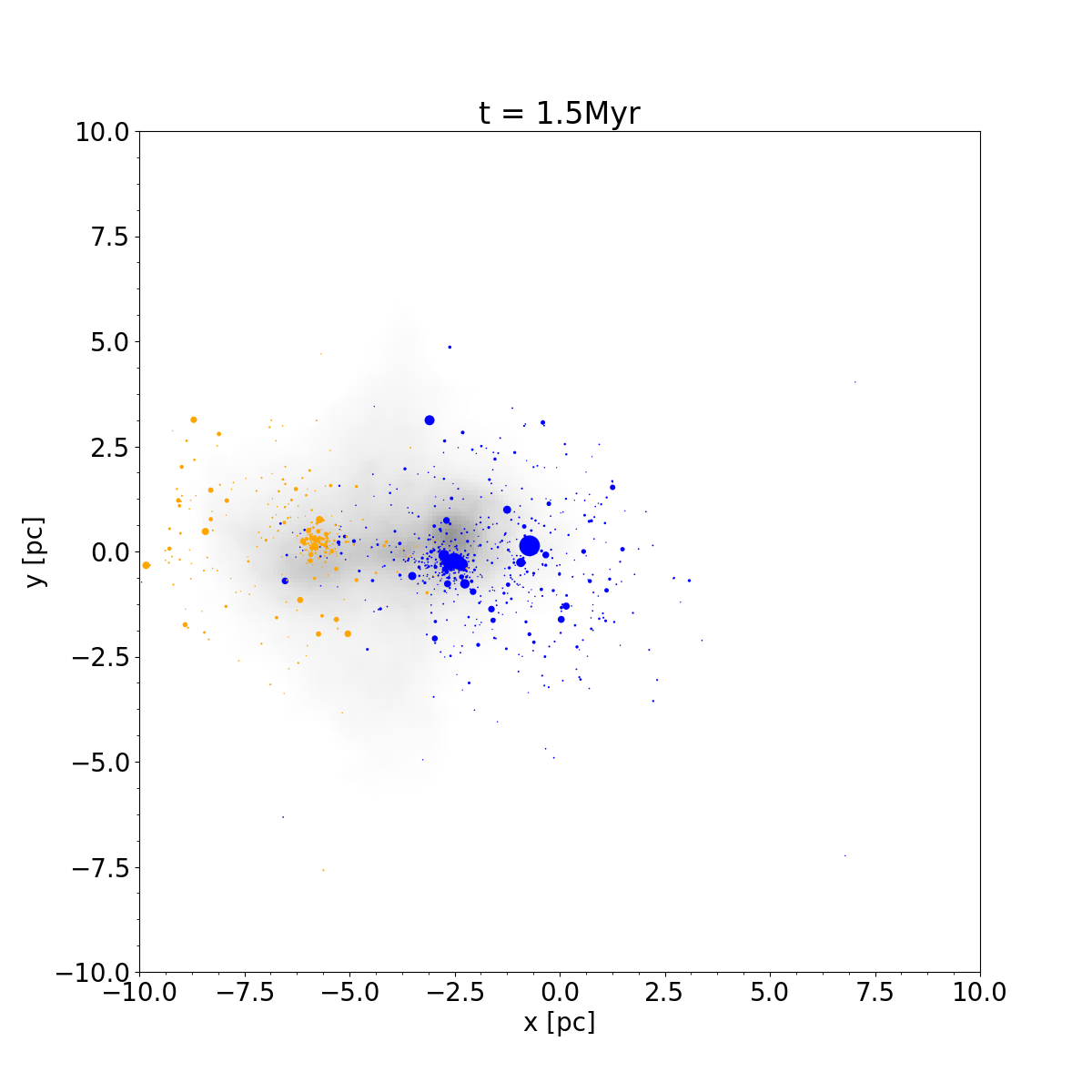}
    \includegraphics[scale=0.24]{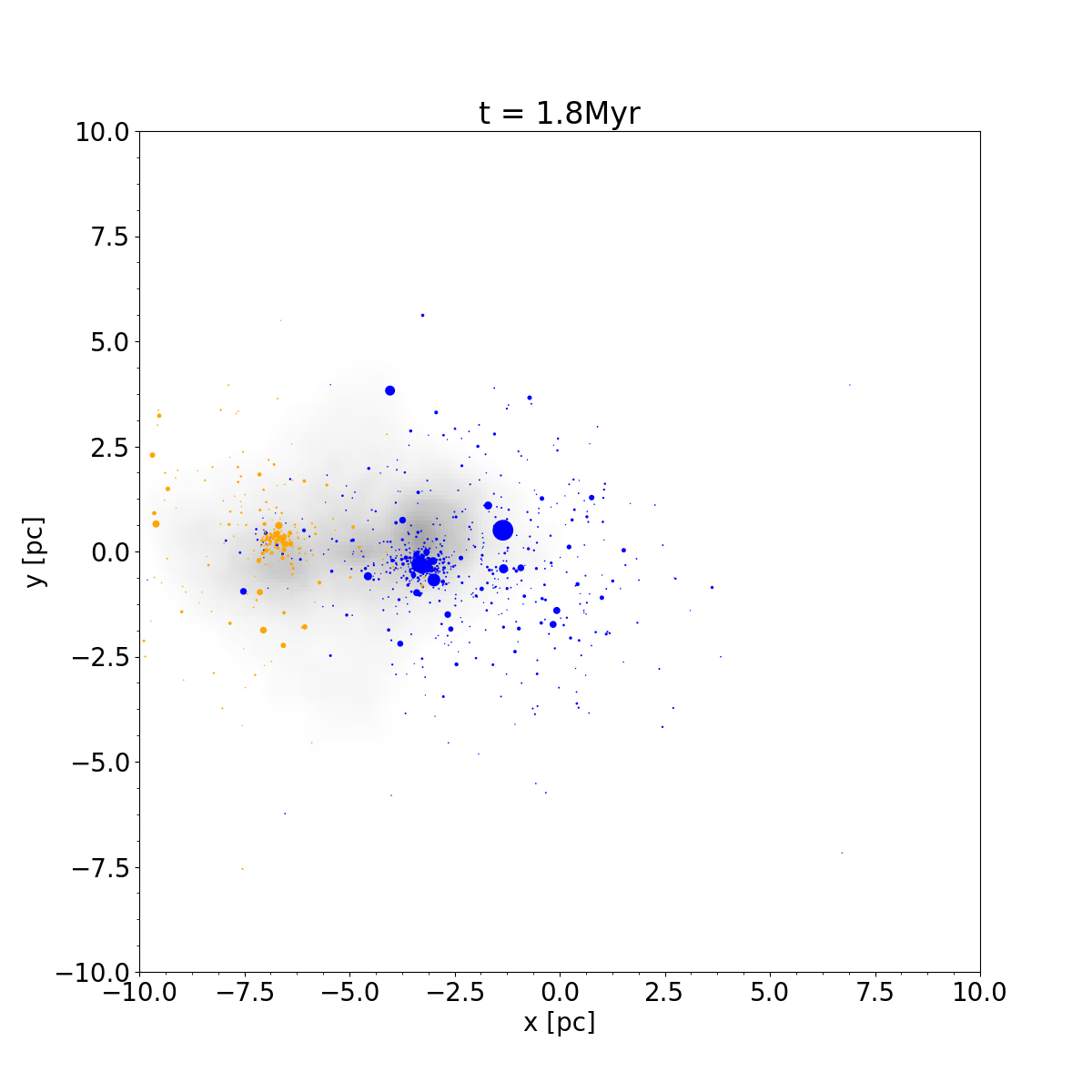}
    \caption{Same as figure \ref{fig:model1} but for run2\_2.8v, one of our non-monolithic simulations. The collision time for this simulation occurs at $t = 0.3$ Myr}
    \label{fig:non-mono}
\end{figure*}

We now briefly discuss the simulations where the two clusters do not form a single entity by the end of our simulation. An example of such a simulation can be seen in figure \ref{fig:non-mono}.

The three simulations that satisfy this criteria are run2\_2p8v, run5, and run4\_2p5v (table \ref{tab:collisions}). They all have collisional velocities $\gtrsim$ 10kms$^{-1}$ and have fairly low mass ratios $f_M \lesssim 5$. The simulations have different final properties -- one of them forms a single gas cloud but not a single stellar cluster, and the other two both continue to show two separate gas clouds and two stellar components.

We find that these three simulations lose much more bound stellar and gas mass than the monolithic simulations. We find that the change in stellar core and half-mass radii for the non-monolithic simulations does not follow the same trend as the monolithic simulations (see figure \ref{fig:inner}, where the 'x' symbols show the final properties of the individual stellar clusters compared to their initial values). Most importantly, however, is the fact that these collisions do not result in mergers, in contradiction to the assumption in H18. One of these collisions is a first-contact merger from H18 (run5) so the fact that it does not, in fact, result in a single monolithic cluster even after 3 Myr may have significant implications for the build-up of massive clusters.

\section{Summary and Discussion}
\label{sec:conc}
We have taken simulation results from \citet{Howard2018} (H18), who found subcluster formation in the filaments of a GMC, and we have resolved their sink particles into detailed models of stars and gas. We have extracted 10 subclusters from the H18 simulations and created detailed models of their stellar and gaseous distribution. We ran 15 merger simulations using these clusters. We found that all resultant clusters lose a small fraction of their bound stars and gas after the merger, but that this fraction increases significantly when we simulate mergers with collisional velocities $\gtrsim 10$kms$^{-1}$. We find that the outermost regions of most of our clusters grow continuously by the end of our simulations, suggesting that the sink particles may be artificially small. We also found that the merger process decreases the amount of gas above star forming densities in the cluster, suggesting that the sink particle prescription may be over-forming stars. We also found that some of our collisions did not result in monolithic clusters, and that others took a long time to form a single cluster. 

The sink particle size and its affect on the accretion of surrounding gas onto the sink particle has been studied in larger scale galaxy and GMC simulations. One example is the converging galactic flows simulation performed by \citet{dobbs_pettitt} who find that through increasing their accretion radius by a factor of $\approx 2.5$, the mass attainable by a sink particle can increase by one order of magnitude. It may be necessary for the treatment of merging sink particles to include something like a "radius of influence" that can change with time in the simulation, and more accurately represent the size of the population inside the sink. 

Our suite of simulations outlines the region of parameter space most occupied by first contact mergers (see figure \ref{fig:parameter_space}). More simulations can help strengthen trends we see in our results. In particular, it would be helpful to model the higher velocity regime where second contact or higher mergers take place. A larger sample size in this regime can help us better understand the conditions for creating a single stellar cluster after a collision. Furthermore, the degree of importance of non-monolithic mergers in GMC simulations may be a byproduct of of the initial conditions of that simulation. For example, \citet{H16} find that velocities of sink particles can be much larger in simulations with lower initial virial parameter (more bound) due to stronger two body interactions. As well, our simulations only involve clusters on the low mass end of those formed in the H18 simulations. The response of higher mass clusters to the merger process is key in understanding the entire formation of the final clusters. We also model all of our subcluster mergers as head on  collisions. We know that off-axis collisions can result in significantly different structures of the collision product \citep[e.g.]{Sills2001}. We will address many of these questions in future work.

We have treated the initial conditions of the stellar population in a fairly simplistic way -- a Plummer sphere, no stellar evolution, and we do not include binary stars. Binaries have been shown to increase the number of high velocity, unbound stars (runaway stars) around star clusters (\citealt{leanard}) and provide an energy source that can heat clusters. \citet{fujiizwart2012} show that the cluster merger process increases the number of newly formed binaries thus resulting in an increase in runaway stars within 3 Myr. The authors find an average of $\approx$ 5 of the cluster's stars become runaway stars at 3 Myr. The inclusion of a binary star prescription would therefore only slightly increase the unbound stellar mass percentage in our simulations showing that our simulations provide a lower limit to the unbound stellar mass that is the result of mergers.

Our simulations of mergers of gas-rich young stellar clusters have shown that they are far from simple. This work provides some hints of the complexity of processes and outcomes that can occur during the build-up of star clusters during their formation. Our future work will investigate these areas further, with the ultimate goal of following cluster merger trees from simulations like H18, and building up young clusters in detail. 

\section*{Acknowledgements}

The authors thank Claude Cournoyer-Cloutier, Michelle Nguyen, Marta Reina-Campos, Veronika Dornan, and Aiyman Mourad for helpful discussions. The authors also thank the anonymous referee for careful reading and constructive comments. AS is supported by the Natural Sciences and Engineering Research Council of Canada. This research was enabled in part by support provided by Compute Ontario (https://www.computeontario.ca) and Compute Canada (http://www.computecanada.ca).

\section*{Data Availability}

The simulations described in this paper are available upon reasonable request to the corresponding author.

\bibliographystyle{mnras}
\bibliography{references} 


\bsp	
\label{lastpage}
\end{document}